\documentstyle[emulateapj]{article}
\newcommand{\ts}{\thinspace}
\begin{document}

\title{%
  DISCOVERY OF A LOW SURFACE BRIGHTNESS OBJECT
  NEAR SEYFERT'S SEXTET
}

\author{\sc
  Takashi Murayama\altaffilmark{1},
  Shingo Nishiura,
  and
  Tohru Nagao\altaffilmark{1}
}

\affil{%
  Astronomical Institute, Graduate School of Science,
  Tohoku University, Aoba, Sendai 980-8578, Japan;
  murayama@astr.tohoku.ac.jp, nishiura@astr.tohoku.ac.jp,
  tohru@astr.tohoku.ac.jp
}

\author{\sc
  Yasunori Sato\altaffilmark{1}
}

\affil{%
  Institute of Space and Astronautical Science,
  3-1-1 Yoshinodai, Sagamihara, Kanagawa 229-8510, Japan;
  sato@astro.isas.ac.jp
}
  
\author{\sc
  Yoshiaki Taniguchi\altaffilmark{1}
}

\affil{
  Astronomical Institute, Graduate School of Science,
  Tohoku University, Aoba, Sendai 980-8578, Japan;
  tani@astr.tohoku.ac.jp
}

\author{\sc and}

\author{\sc
  D.~B.~Sanders
}

\affil{%
  Institute for Astronomy, University of Hawaii,
  2680 Woodlawn Drive, Honolulu, HI 96822;
  sanders@ifa.hawaii.edu
}

\altaffiltext{1}{%
  Visiting Astronomer of the University of Hawaii
  2.2 meter telescope.
}

\authoremail{murayama@astr.tohoku.ac.jp}

\begin{abstract}
We report the discovery of a low surface brightness
(LSB) object serendipitously found during deep CCD imaging of
a compact group of galaxies, Seyfert's Sextet, 
in $V\!R$ and $I$ bands.
The LSB object is located 2\farcm3 southwest from
the group center.
Its surface brightness within the angular effective radii of
$r_{\rm e}(V\!R)=4\farcs8$ and
$r_{\rm e}(I)=4\farcs8$ is very low -- 
$\mu_{\rm e} (V\!R)=25.28$ mag arcsec$^{-2}$ and
$\mu_{\rm e}(I)=24.47$ mag arcsec$^{-2}$, respectively.
The apparent magnitudes are $m_{\rm AB}(V\!R)=19.87$ mag and
$m(I)=19.06$ mag.
The object is most likely a LSB dwarf galaxy, but 
other possibilities are also discussed.
\end{abstract}


\section{INTRODUCTION}
Low surface brightness (LSB) galaxies 
have been extensively discussed within the context of the 
overall formation and evolution of galaxies as well as 
observational cosmology, and in particular for their possible contribution
as a major fraction of the total galaxy population
(see \markcite{Imp97}Impey \& Bothun 1997, for a review).
During deep imaging observations of the compact
group of galaxies known as Seyert's Sextet
(\markcite{Nis99}Nishiura et al.\ 1999)
we found a LSB galaxy candidate near the group.
In the present paper, we report on 
the photometric properties of this LSB candidate.
We adopt a Hubble constant of 100 $h$ km s$^{-1}$
Mpc$^{-1}$ throughout this paper.

\section{OBSERVATIONS AND DATA REDUCTION}
The observations were carried out
at the University of Hawaii 2.2{\ts}m telescope
using the 8192$\times$8192 (8k) CCD Mosaic camera
(\markcite{Lup96}Luppino et al.\ 1996).
The camera was attached at the f/10 Cassegrain focus
and provided a $\sim 18\arcmin\times18\arcmin$ field of view.
The CCDs were read out in the $2 \times 2$ pixel binning mode
which gave an image scale of 0\farcs26 pixel$^{-1}$.
We obtained broad band images with the $V\!R$ and $I$ filters on
1999 May 20 and May 23 (UT), respectively.
The integration time for each exposure was set to 8 minutes.
Twenty-three exposures for the $V\!R$-band and 24 exposures for the $I$-band 
were taken, thus the total integration time amounted
to 184 minutes in $V\!R$ and 192 minutes in $I$.

Data processing was done in a standard way
using IRAF\footnote{%
Image Reduction and Analysis Facility (IRAF) is distributed 
by the National Optical Astronomy Observatories, which are operated 
by the Association of Universities for Research in Astronomy, 
Inc., under cooperative agreement with the National Science
Foundation.}.
After bias and dark counts were subtracted,
each frame was divided by the flatfield image,
which was a median image of all of the object frames
obtained during a night.
The object frames were median-combined with their
positions registered.
Typical seeing, as estimated from the processed images, 
was $\sim${\ts}0\farcs8 in both bands.
Standard stars from Landolt (1992\markcite{Lan92}) were observed
and used for calibration of absolute fluxes.
Since the $V\!R$ filter is not a standard photometric band
(see \markcite{Jew96}Jewitt, Luu, \& Chen 1996),
we adopted an AB magnitude scale for this bandpass.
The absolute photometric errors were estimated to be 
$\pm 0.05$ mag for the $V\!R$-band and
$\pm 0.03$ mag for $I$-band. 
The limiting surface brightnesses are 
$\mu_{V\!R}^{\rm lim}=28.7$ mag arcsec$^{-2}$ and
$\mu_{I}^{\rm lim}=28.1$ mag arcsec$^{-2}$, corresponding to
a 1$\sigma$ variation in the background.

\section{RESULTS}
In Figure \ref{ssimage}
we show the $V\!R$- and $I$-band images
of Seyfert's Sextet.
A faint, extended object is 
located in both bands at 2\farcm3 southwest
from the group center of Seyfert's Sextet.
Our estimate of the centroid position of this faint object is 
$\alpha$(B1950)=15$^{\rm h}$ 56$^{\rm m}$  51\fs6,
$\delta$(B1950)=+20\arcdeg{} 52\arcmin{} 42\arcsec.
As shown in the lower panels of Figure \ref{ssimage},
the shape of the object appears to be nearly spherical in both bands
and very diffuse compared with the 
foreground/background galaxies around it.
There is no evidence that the object moved during
the $V\!R$ and $I$ observations
(the $V\!R$ image was taken three days prior to the $I$ band image),
thus the object is not likely to be within our solar-system.

\placefigure{ssimage}

Using a wavelet package in ESO-MIDAS,\footnote{%
European Southern Observatory Munich Image Data Analysis System
(ESO-MIDAS) is developed and maintained by
the European Southern Observatory.}
we applied a Wiener-like wavelet filter
to the images in order to improve the signal-to-noise. 
The results are shown in Figure \ref{lsbimage}.
In the central region of the processed $V\!R$ image
there seem to be two intensity peaks lying along
a northwest-southeast direction
with a separation of $\sim 2\farcs5$.
On the other hand, the $I$-band image shows a single
peak which is located between the two $V\!R$ peaks.
This could be interpreted as being due to an inhomogeneous distribution of 
dust, or perhaps a strong emission line from ionized gas that
appears in either band.
Alternatively, the intensity peaks near the center
may be background galaxies.
At 19\arcsec{} west of the center of the faint object
there is another diffuse condensation which
could perhaps be interpreted as a tidal structure.
This companion structure appears to be present 
in the original images, but given that the strength of this 
feature is comparable
to the noise uncertainty in the original images,
we will not discuss it further in this paper.

\placefigure{lsbimage}

Next we discuss the photometric properties of this LSB object.
Since the object appears to be nearly spherical, we will adopt a 
circular aperture for computing the radial light distribution.
Apparent foreground/background objects were first masked, 
and the surface brightness on the original images
was then determined using 0\farcs8 radial bins where the binwidth was set 
to be approximately equal to the seeing.
The center of the aperture was fixed at the position of 
the intensity peak in the noise reduced $I$-band image.
Figure \ref{profile} shows the surface brightness profiles
in both bands.
These profiles cannot be simply fit with an exponential disk
as can be seen by the fact that an exponential profile would appear as a straight line
on the $\mu$-$r$ plot (the upper panel of Figure \ref{profile}).
Although the observed radial profiles could be approximated by a 
straight line fit at radii $4\arcsec{} \lesssim  r \lesssim 10\arcsec$, 
it is clear that  the profiles flatten 
at $r\lesssim 4\arcsec$.
Furthermore, a $r^{1/4}$-law profile, which is more centrally concentrated
than an exponential profile, is also a poor approximation to 
the observed data points.
We also note that since the flattened region of the profiles is large
compared to the seeing size of $\sim 0\farcs8$,
the flat profiles at $r \lesssim 4\arcsec$ are a genuine property of this LSB object.   

In order to more accurately approximate the observed radial surface brightness distribution, 
we decided to adopt a $r^{1/n}$-law fit with $n<1$;
\[
\mu(r) = \mu_0 + 2.5 (\log_{10} e) \left(\frac{r}{s}\right)^{1/n},
\]
where $\mu(r)$ is the surface brightness at a radius of
$r$ from the center, $\mu_0$ is the central surface brightness, and
$s$ is the angular scale length.
This profile is less concentrated than an exponential profile, i.e.,
shallower at the inner area and a steeper
profile at the outer area than an exponential profile.
The solid curves shown in Figure \ref{profile}
are the best fit for each band.
We used only the points at $1\arcsec < r < 10\arcsec$
(shown by the filled circles in Figure \ref{profile})
to avoid the seeing effect at the center
and the sky-noise limited area at large radii.
Table \ref{nfit} lists the fit parameters 
($n$, $\mu_0$, and $s$) as well as other photometric properties
which were subsequently derived.
Our analysis shows that the LSB galaxy has a $r^{1/n}$ surface
brightness profile with $n\sim0.6$. 

\begin{deluxetable}{lcc}
\tablenum{1}
\tablecaption{%
 Results of $r^{1/n}$-law Fitting and Photometric Properties.\tablenotemark{a}
 \label{nfit}}
\tablehead{
  &
  \colhead{$V\!R$} &
  \colhead{$I$}
}
\startdata
Exponent of $r^{1/n}$ law ($n$)    & $0.57 \pm 0.01$ & $0.65 \pm 0.01$ \nl
Central surface brightness ($\mu_0$)  [mag arcsec$^{-2}$]
                                   & $24.83 \pm 0.02$ & $23.90\pm 0.03$ \nl
Angular scale length ($s$) [arcsec]
                                   & $5.35 \pm 0.07$ & $4.84 \pm 0.08$ \nl
\hline
Total apparent magnitude ($m$) [mag]
                                   & $19.87 \pm 0.02$ & $19.06 \pm 0.03$ \nl 
Angular effective radius ($r_{\rm e}$) [arcsec]
                                   & $4.81\pm 0.07$ &  $4.81 \pm 0.08$ \nl
Effective surface brightness ($\mu_{\rm e}$) [mag arcsec$^{-2}$]
                                   & $25.28 \pm 0.02$ & $24.47 \pm 0.03$ 
\enddata
\tablenotetext{a}{Errors quoted in this table are formal errors
in the fitting.}
\end{deluxetable}

\placefigure{profile}

\section{DISCUSSION}
We first discuss the observed properties of our candidate LSB galaxy in terms of 
the known properties of LSB galaxies. 
The color of our LSB candidate is $V\!R-I \simeq 0.81$.
Since the $V\!R$-band is inconvenient for comparison with standard photometry of galaxies,
we first estimated that $V-I \simeq 0.94$ and $R-I \simeq 0.49$ by
interpolating the observed fluxes in the $V\!R$-band and $I$-band.
These colors are not peculiar for LSB dwarfs or LSB disk galaxies
(e.g., \markcite{Imp97}Impey \& Bothun 1997).
One of the characteristics of our LSB object is
the $r^{1/n}$ surface brightness profile with $n \sim 0.6$, however 
the majority of LSB galaxies exhibit an exponential profile, i.e.,
$n=1$ (\markcite{One97}O'Neil, Bothun, \& Cornell 1997).
On the other hand, many of the faiter dwarf ellipticals in the
Fornax cluster have less concentrated profiles than the exponential
(\markcite{Cal87}Caldwell \& Bothun 1987).
\markcite{Dav88}Davies et al.\ 1988 also showed that
a significant number of LSB dwarf galaxies
in the Fornax cluster show $r^{1/n}$-law profiles with $n<1$.
Furthermore, \markcite{One97}O'Neil et al.\ (1997)
showed that 17{\ts}\% of the LSB galaxies in their sample have
less concentrated surface brightness profiles
than an exponential disk although they did not fit the profiles
with the $r^{1/n}$ law but instead used a King model profile.

\markcite{Cao93}Caon, Capaccioli, \& D'Onofrio (1993)
found that the value of $n$ is well correlated with the effective
radius ($R_{\rm e}$)
for spheroidal galaxies ranging from the LSB dwarfs in
the sample of \markcite{Dav88}Davies et al.\ (1988) to 
the giant ellipticals in the Virgo cluster.  In this context, the observed 
exponent of $n \sim 0.6$ for our LSB candidate 
implies that this object is indeed a good candidate for a LSB dwarf galaxy.
As shown in Figure 5 of \markcite{Cao93}Caon et al.\ (1993),
the LSB dwarfs with $n<1$ have $R_{\rm e}$ in a range between
$\simeq 0.13$ kpc and $\simeq 1.3$ kpc.
Given the angular effective radius of our LSB galaxy candidate 
of $r_{\rm e}=4\farcs8$, this would imply a distance somewhere in the range
of 5.4 Mpc to 54 Mpc, and a corresponding absolute $I$-band magnitude of
between $-10$ mag and $-15$ mag,  
which is comparable to those of dwarf galaxies in the Local Group
(\markcite{Mat98}Mateo 1998).
This would imply that our LSB galaxy candidate would be at the faint end of
the luminosity function of LSB galaxies
(e.g. \markcite{Imp97}Impey \& Bothun 1997).

If the LSB galaxy is located at the same distance
as Seyfert's Sextet (44 $h^{-1}$ Mpc),
one might conclude that the LSB galaxy may have been formed
through possible tidal interactions
between the group galaxies.
The projected separation of the LSB galaxy from
the group center is $\approx 30$ $h^{-1}$ kpc.
The LSB galaxy could travel this distance
in $2 \times 10^8$ years assuming a a projected velocity equal to 
the radial velocity dispersion (138 km s$^{-1}$)
of the group.

At smaller distances than that of Seyfert's Sextet,
only one galaxy is known within 1\arcdeg{} 
of the LSB candidate. 
It is also a LSB galaxy, F583-1 (= D584-04: 
\markcite{Sch88}Schombert \& Bothun 1988;
\markcite{Sch92}Schombert et al.\ 1992;
\markcite{Sch97}Schombert, Pildis, \& Eder 1997).
The distance toward F583-1 corresponding to its redshift
is 25 $h^{-1}$ Mpc.
If our LSB galaxy is located at the same distance as F583-1,
the apparent separation of 23\arcmin{} between
F583-1 and the LSB galaxy corresponds to 170 $h^{-1}$ kpc.

Another possibility is that the LSB object
is located at a much smaller distance.
This would imply that the LSB object may be a system more like
Galactic globular clusters.
In fact, a King model profile with a concentration parameter
of $\sim 0.7$ and a core radius of $\sim 4\farcs{}0$ is also a food fit to 
the present LSB object (Note: we have not shown this fit since it is 
very similar to that shown in Fig.\ 3).
\markcite{One97}O'Neil et al.\ (1997) noted a possibility that some
LSB objects well fitted with the King profile that were found in their survey
may be Galactic LSB globular clusters.
Although the concentration parameter of $c \sim 0.7$ of our LSB object
is smaller than those of typical Galactic globular clusters
(e.g., \markcite{Chr89}Chernoff \& Djorgovski 1989),
globular clusters in the outer halo of the Galaxy
($\sim${\ts}30--100 kpc from the Galactic center)
have concentration parameters as small as our LSB object
(see, for example, \markcite{Djo94}Djorgovski \& Meylan 1994).
The central surface brightness of the globular clusters 
in the Galactic halo can be as faint as
$\sim 24$ mag arcsec$^{-2}$ in $V$,
which is comparable to those of LSB galaxies.
The clusters in the Galactic halo have larger core radii of $\sim 20$ pc
than globular clusters at smaller galactocentric radii because of smaller tidal forces at
larger distance from the galactic center.
If the present LSB object is such a distant globular cluster,
it should have a core radius of $\sim 20$ pc.
Thus, the apparent core radius of $r_{\rm c} \sim 4\farcs0$
leads to a distance toward the LSB object of  $\sim 1$ Mpc.
However if this is the case, stars in the LSB would be resolved
spatially in our images. Therefore, this possibility can be rejected.

In summary, we have discovered a LSB object near the compact group of
galaxies known as Seyfert's Sextet.
The LSB object is likely to be one of the following:
1) a field LSB dwarf galaxy at a distance of 5.4--54 Mpc, or 
2) a LSB dwarf galaxy at the same distance of Seyfert's Sextet
(44 $h^{-1}$ Mpc).
Measurement of the redshift, either by optical spectroscopy
or by radio observations of \ion{H}{1} gas
will be necessary to determine which of these descriptions applies.

\acknowledgments
The authors are very grateful to the staff of the UH 2.2 m telescope.
In particular, we would like to thank
Andrew Pickles for his technical support and assistance during
the observations.
We also thank Richard Wainscoat and Shinki Oyabu for their kind help on 
photometric calibration,
Tadashi Okazaki for kindly providing us
his program for calculating King model profiles, 
and Daisuke Kawata for helpful comments.
This work was financially supported in part by Grants-in-Aid for 
Scientific Research (Nos. 07055044, 10044052, and 10304013)
from the Japanese Ministry of Education, Science, Sports, and Culture
by the Foundation for Promotion of Astronomy, Japan.
TM is thankful for support from a Research Fellowship from the Japan
Society for the Promotion of Science for Young Scientists.
This research has made use of the NASA/IPAC Extragalactic Database
(NED) and the NASA Astrophysics Data System Abstract Service.

\clearpage

\figcaption[fig1.ps]{%
The $V\!R$- and $I$-band images of Seyfert's Sextet (upper panels).
The newly discovered LSB object is in the center of the box at the lower right.
\label{ssimage}
}

\figcaption[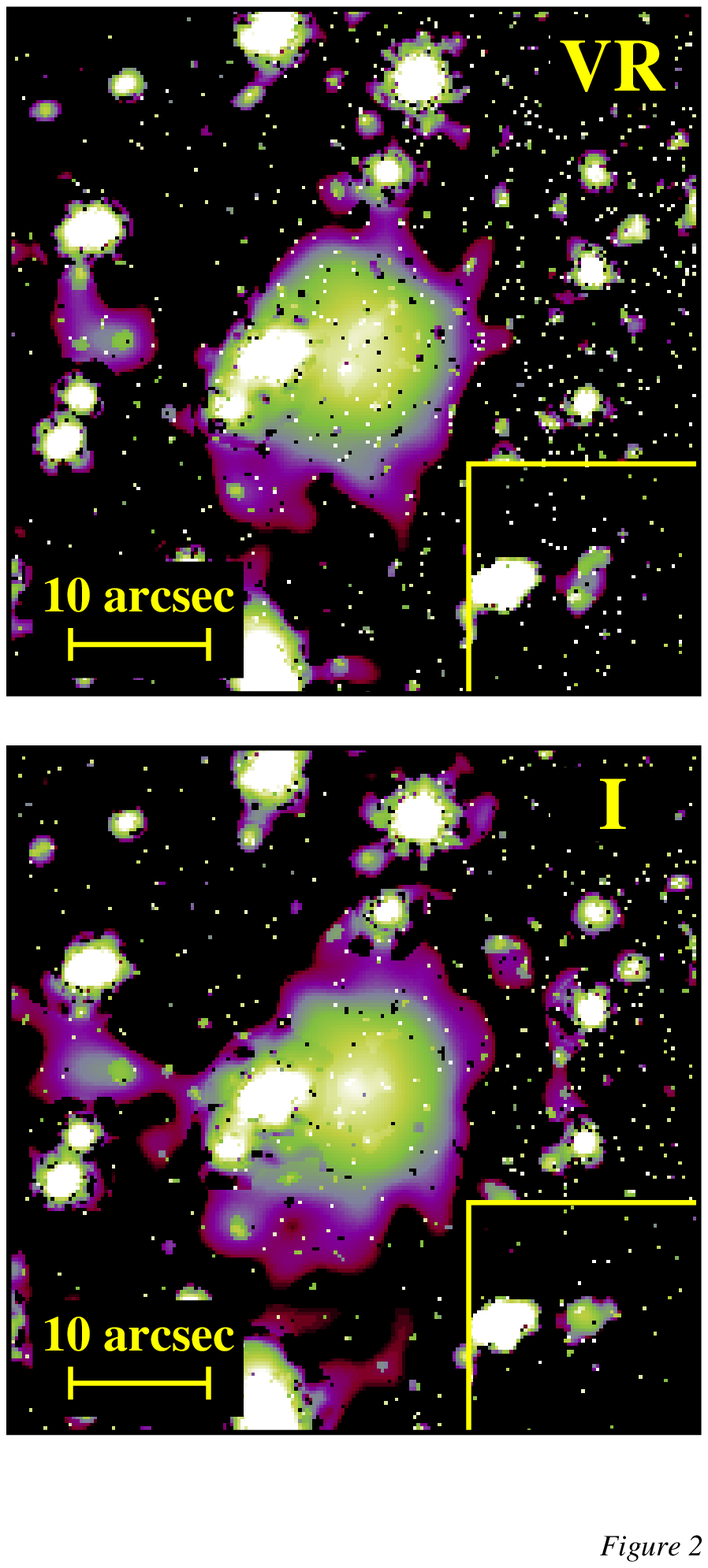]{%
The noise reduced images of the LSB object
processed with a Wiener-like wavelet filter.
In the lower panels the boxed region is shown at a 
different intensity scale.
\label{lsbimage}
}

\figcaption[fig3.ps]{%
The surface brightness profiles of the LSB object
in the $V\!R$- and $I$-bands versus 
the angular radius in linear scale (upper panel) and
in logarithmic scale (lower panel).  
Observed data are shown as filled and open circles with
error bars.
The filled circles denote the data points which were used
for the $r^{1/n}$-law fit.
The results of the $r^{1/n}$-law fit are shown by
the solid lines.
\label{profile}
}

\end{document}